\documentclass[aps,prl,twocolumn,superscriptaddress,a4paper]{revtex4}

\usepackage{graphicx}
\usepackage{xfrac}

\pdfoutput=1 

\usepackage{fancyhdr}
\usepackage[colorlinks=true,citecolor=blue,linkcolor=magenta]{hyperref}
\hypersetup{%
pdfauthor = {Patrick Maletinsky},
colorlinks = true, linkcolor = blue, urlcolor=blue, bookmarksnumbered =  true}

\begin{document}


\title{Sub-nanometer resolution in three-dimensional magnetic-resonance imaging of individual dark spins}


\author{M.S. Grinolds}
\affiliation{Department of Physics, Harvard University, Cambridge, Massachusetts 02138 USA}
\author{M. Warner}
\affiliation{Department of Physics, Harvard University, Cambridge, Massachusetts 02138 USA}
\author{K. De Greve}
\affiliation{Department of Physics, Harvard University, Cambridge, Massachusetts 02138 USA}
\author{Y. Dovzhenko}
\affiliation{Department of Physics, Harvard University, Cambridge, Massachusetts 02138 USA}
\author{L. Thiel}
\affiliation{Department of Physics, University of Basel, Klingelbergstrasse 82, Basel CH-4056, Switzerland}
\author{R.L. Walsworth}
\affiliation{Harvard-Smithsonian Center for Astrophysics, Cambridge, Massachusetts 02138 USA}
\author{S. Hong}
\affiliation{Vienna Center for Quantum Science and Technology (VCQ), Faculty of Physics, University of Vienna, A-1090 Vienna, Austria.}
\author{P. Maletinsky}
\affiliation{Department of Physics, University of Basel, Klingelbergstrasse 82, Basel CH-4056, Switzerland}
\author{A. Yacoby}
\affiliation{Department of Physics, Harvard University, Cambridge, Massachusetts 02138 USA}
\email[]{yacoby@physics.harvard.edu}

\date{\today}

\begin{abstract}
Magnetic resonance imaging (MRI) has revolutionized biomedical science by providing non-invasive, three-dimensional biological imaging\,\cite{Mansfield2004}. However, spatial resolution in conventional MRI systems is limited to tens of microns\,\cite{Glover2002}, which is insufficient for imaging on molecular and atomic scales. Here we demonstrate an MRI technique that provides sub-nanometer spatial resolution in three dimensions, with single electron-spin sensitivity. Our imaging method works under ambient conditions and can measure ubiquitous 'dark' spins, which constitute nearly all spin targets of interest and cannot otherwise be individually detected. In this technique, the magnetic quantum-projection noise of dark spins is measured using a single nitrogen-vacancy (NV) magnetometer located near the surface of a diamond chip. The spatial distribution of spins surrounding the NV magnetometer is imaged with a scanning magnetic-field gradient. To evaluate the performance of the NV-MRI technique, we image the three-dimensional landscape of dark electronic spins at and just below the diamond surface and achieve an unprecedented combination of resolution (0.8 nm laterally and 1.5 nm vertically) and single-spin sensitivity. Our measurements uncover previously unidentified electronic spins on the diamond surface, which can potentially be used as resources for improved magnetic imaging of samples proximal to the NV-diamond sensor. This three-dimensional NV-MRI technique is immediately applicable to diverse systems including imaging spin chains, readout of individual spin-based quantum bits, and determining the precise location of spin labels in biological systems.
\end{abstract}

\maketitle

Atomic-scale magnetic resonance imaging (MRI) would have wide-ranging applications including determining the structure of individual biomolecules\,\cite{Hemmer2013}, imaging the dynamics of bottom-up molecular engineering\,\cite{Palma2011}, and achieving site-resolved readout in solid-state quantum simulators\,\cite{Cai2013}. Performing conventional MRI on sub-micron length scales is not possible because macroscopically generated magnetic-field gradients limit spatial resolution, and inductive detection schemes suffer from significant thermal noise\,\cite{Glover2002}. Great progress has been made using scanning-probe-based magnetic gradient techniques, which enable nanoscale MRI,\cite{Sidles1995,Balasubramanian2008} using ultrasensitive force detection at cryogenic temperatures\,\cite{Degen2009,Rugar2004} or fluorescence measurements of optically 'bright' spins such as nitrogen vacancy (NV) color centers in diamond\,\cite{Balasubramanian2008,Grinolds2011}. However for most MRI applications, measurements must be taken near room temperature, and nearly all targets of interest contain optically 'dark' spins that are unpolarized or weakly polarized. In this work we demonstrate a technique to perform three-dimensional (3D) MRI with sub-nanometer resolution on dark electronic spins under ambient conditions, using a single NV center near the surface of a diamond as a magnetic sensor of its local environment, together with a scanning-tip magnetic-field gradient to provide high spatial resolution. Our method is compatible with numerous developed methods for bringing imaging targets sufficiently close for NV magnetic detection\,\cite{Maletinsky2012,Grotz2011,Mamin2013,Mamin2012,Staudacher2013}, and extends the reach of nanoscale MRI to previously inaccessible systems in both the physical and life sciences.

\begin{figure*}
\includegraphics[scale=1]{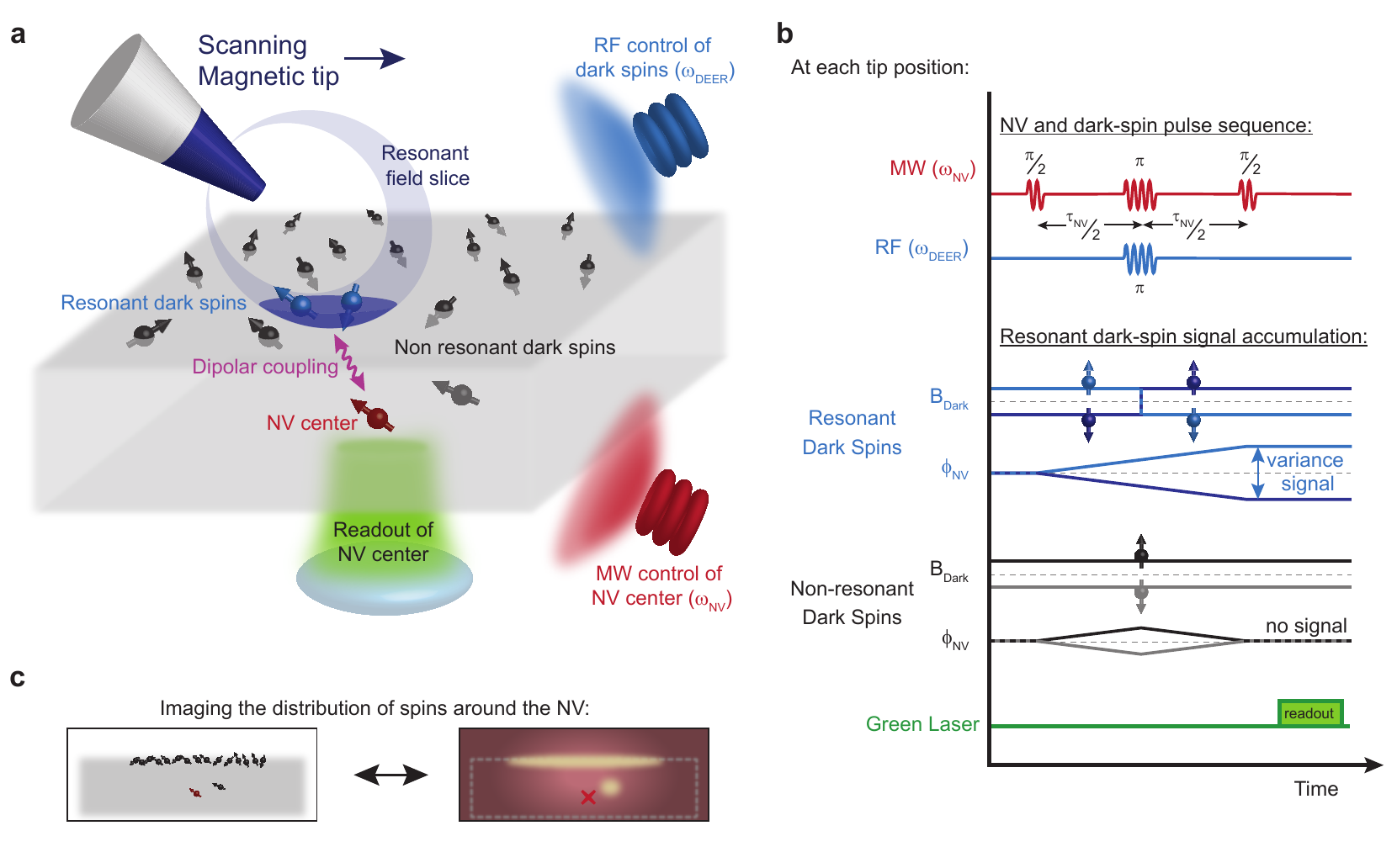}
\caption{\label{Fig1} Dark-spin MRI using scanning gradients and a single nitrogen-vacancy sensor. {\bf a} Schematic of NV-MRI technique depicting an NV center in diamond situated in a confocal laser spot with nearby dark spins. A scanning magnetic tip is placed within 100 nm of the diamond surface. Applied microwave (MW) and radio frequency (RF) signals allow for independent coherent control of the NV spin and dark spins. By scanning the magnetic tip, non-resonant dark spins (shown in black) are systematically brought into resonance with the RF signal (resonant spins shown in blue)  and are measured via optically detected magnetic resonance of the NV sensor. {\bf b} Double electron-electron resonance (DEER) pulse sequences executed at each magnetic tip position. A MW spin-echo sequence is executed on the NV sensor. By synchronizing an RF $\pi$ pulse on the dark spins with the MW $\pi$ pulse in the echo sequence, the time-varying magnetic field from the dark spins (BDark) in the resonant slice (light/dark blue) leads to net NV spin phase ($\tau_{NV}$) accumulation, while the magnetic field from non-resonant dark (grey/black) spins is refocused and thus their effects on the NV spin are cancelled, irrespective of the initial polarization state of the dark spins. {\bf c} NV-MRI provides 3D mapping of the distribution of dark electronic spins near the NV sensor (indicated by the red cross), with sub-nanometer resolution (see main text for further discussion).}
\end{figure*}

The NV-MRI technique combines an NV magnetometer with scanning magnetic-field gradients using an atomic-force microscope (Fig. \,\ref{Fig1}a). Individual shallowly implanted NV centers (nominal depth of 10 nm) are placed in the focus of a confocal microscope, so that the NV electronic spin can be initialized by optical pumping, used as a sensor to measure nearby dark spins, and read out using time-dependent fluorescence\,\cite{Gruber1997}. To image the 3D distribution of dark spins via NV-MRI, we apply a local magnetic-field gradient with a scanning magnetic tip. The magnetic tip provides a narrow spatial volume (a 'resonant slice') in which dark spins are on resonance with a driving radio-frequency (RF) field. Only dark spins within the resonant slice are RF-driven, and thus contribute, to the dark-spin magnetic signal measured by the NV center. The 3D position of the resonant slice is then controllably scanned throughout the sample with angstrom precision by moving the magnetic tip, allowing high-resolution 3D MRI of target dark spins.

To create 3D magnetic resonance images, the detected NV-MRI signal at each magnetic tip position is made conditional on the resonant RF-driving of target dark spins via double electron-electron resonance (DEER)\,\cite{Grotz2011,Mamin2012,deLange2012,Larsen1993}. As illustrated in Fig. \,\ref{Fig1}b, microwave (MW) pulsing on the NV spin prepares a coherent superposition of NV-spin states with phase $\phi_{NV}$ that evolves with evolution time $\tau_{NV}$ in proportion to the local magnetic field (projected along the NV quantization axis) from the target dark spins ($B_{Dark}$). Halfway through $\tau_{NV}$, simultaneous MW and RF $\pi$-pulses are applied to the NV and target dark spins respectively, so that $\phi_{NV}$ accumulates only for resonant dark spins, and refocuses for off-resonant dark spins. Target dark spins are in an unpolarized mixed state at room temperature, and so across multiple spin measurements, $\textless\phi_{NV}\textgreater$ = 0; however, DEER measures $\cos{\phi_{NV}}$ , which is independent of the dark spins' initial states and consequently measures the variance of the dark spins (coming from magnetic quantum-projection noise).

When scanning the magnetic tip to perform NV-MRI, we simultaneously frequency-lock the applied MW signal to the NV spin resonance\,\cite{Schoenfeld2011}, which keeps the NV sensor active and also measures the tip-induced frequency detuning. The resulting spatial map of the frequency-locked NV signal experimentally determines the point-spread-function (PSF) for dark-spin imaging. Because dark spins are spatially offset from the NV location and/or distributed over a non-zero volume, the observed dark-spin signal as a function of magnetic tip position is offset and/or broadened from the measured PSF, and the dark-spin spatial distribution can be found via deconvolution (Fig. \,\ref{Fig1}c). An important feature of our technique is that by directly measuring the dark-spin PSF there is no reliance on magnetic-field modeling or iterative deconvolution schemes that must be simultaneously solved for both an unknown signal and an unknown PSF.

\begin{figure*}
\includegraphics[scale=1]{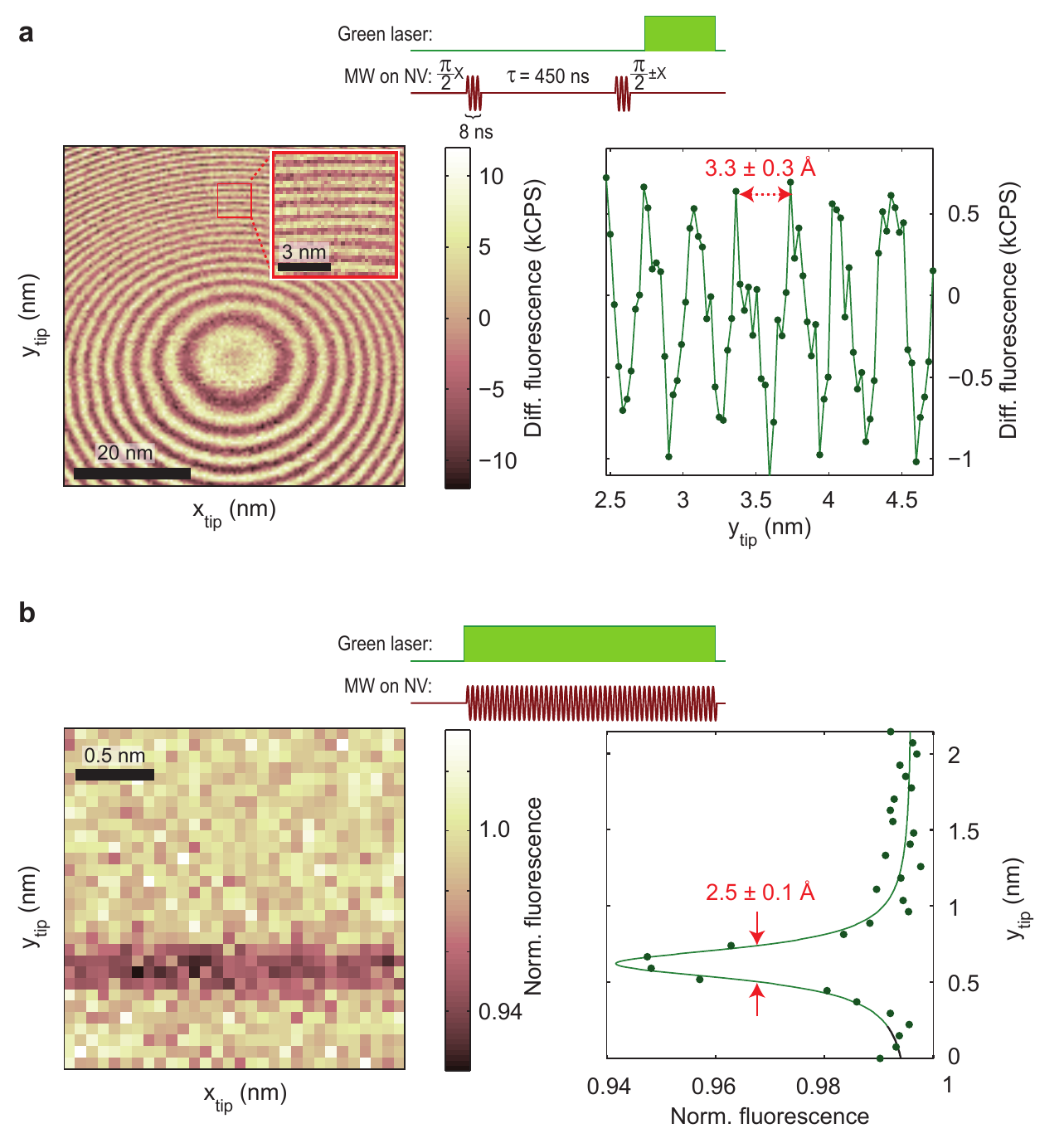}
\caption{\label{Fig2} Scanning gradients with sub-nm MRI resolution. {\bf a} Scanning Ramsey interferometry on a single, shallow NV center in a 0.12-T external magnetic field. As the magnetic tip is laterally scanned over the NV center 50 nm above the diamond surface and near the NV center, the resultant variation in magnetic field at the NV center leads to 2D spatial oscillations in the measured NV Ramsey interferometry fluorescence signal as a function of lateral tip position (center panel). A linecut along the y-direction (30 nm above the diamond surface, right panel) shows oscillations with a period of 3.3 $\pm$ 0.3 $\AA{}$, indicating that the tip-induced magnetic-field gradients are 2.4 G/nm. {\bf b} Continuous-wave ESR magnetometry with an inverse line-width of $\tau$ = 120 ns. Magnetic-tip scan with a 20-nm vertical offset from the NV, zoomed into one resonant slice with a full-width-at-half-maximum of 2.5 $\AA{}$ and thus a 12 G/nm tip-induced magnetic-field gradients.}
\end{figure*}

The spatial resolution of NV-MRI is given by $\frac{1}{\tau\gamma\mid\nabla B_{tip}\mid}$ where $\gamma$ and $\tau$ are the target spin's gyromagnetic ratio and spin-interrogation time, respectively, and $\nabla B_{tip}$ is the gradient of the tip's magnetic field at the target spin's position projected along the spin's quantization axis. We determined the spatial resolution limit in our setup by measuring $\nabla B_{tip}$ using a single NV center, with a relatively long $T_{2}^{*}$ coherence time, which allows for a long $\tau$ (Fig. 2). Using a $\tau$ = 450 ns Ramsey interferometry sequence, we measured spatial fringes with oscillation periods down to 3.3 $\pm$ 0.3 $\AA{}$ (Fig. \,\ref{Fig2}a), showing that the magnetic tip produces a gradient of 2.4 G/nm and demonstrating that the experimental setup is mechanically stable down to sub-nm length scales. By bringing the tip closer to the NV center (Fig. \,\ref{Fig2}b), we observed gradients of at least 12 G/nm (effective $\tau$ = 120 ns); however, vibrations in our experimental setup currently limit the spatial resolution to 2.5 $\AA{}$. For NV-MRI of dark spins with static tip gradients, the target-spin interrogation time is limited the target spin $T_{2}^{*}$ (~150 ns), enabling sub-nanometer 3D NV-MRI resolution.

To demonstrate such sub-nanometer NV-MRI performance, we spatially mapped the spin environment of individual NV centers near a diamond surface. Shallow NV centers are the mainstay for NV-based sensing\,\cite{Maletinsky2012,Grotz2011,Mamin2012,Mamin2013,Staudacher2013,Grinolds2013,Ofori-Okai2012,Ohno2012} and quantum-information processing\,\cite{Cai2013,Dolde2013}, yet their dominant sources of decoherence have not been identified and localized. In the absence of the magnetic tip, we first used DEER spectroscopy and observed a g=2 dark electronic spin bath coupled to shallow NV spins, consistent with previous measurements that did not determine the origin or spatial distribution of these dark spins\,\cite{Mamin2012}. We measured g=2 electron-spin resonances for more than 60\% of measured NV spins (\textgreater30 centers in three diamond samples). We then used the NV-MRI technique to perform 3D imaging of the spatial distribution of these dark electronic spins on and near the diamond surface.

\begin{figure*}
\includegraphics[scale=1]{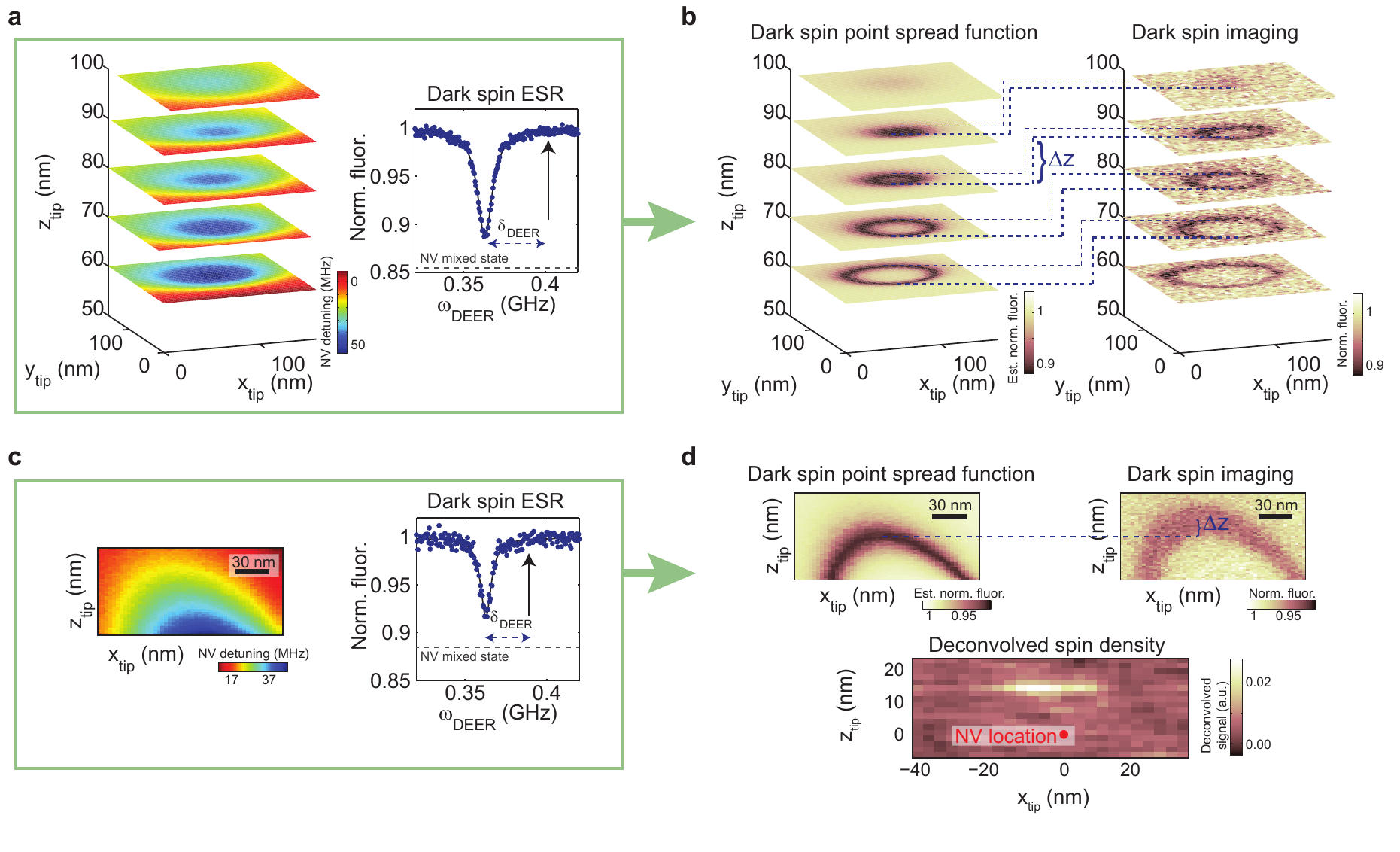}
\caption{\label{Fig3} MRI of ensembles of dark spins at the diamond surface. {\bf a} Determining the dark-spin NV-MRI point-spread function (PSF). 3D plot of the detuning of the frequency-locked NV microwave signal $\omega_{NV}$, measured by xy scanning of the magnetic tip with variable z offset of 60 nm to 100 nm from the diamond surface (left). This detuning map is combined with the tip-independent dark-spin electron-spin resonance (ESR) spectrum (right) to determine the dark-spin PSF. {\bf b} NV-MRI tomography of dark spins proximal to an individual shallow NV center. Displacement of the dark-spin resonance slice image (right) from the dark-spin PSF (left) indicates the location of the dark spins with respect to the NV center. For a given dark-spin lateral (xy) image, the best match to the dark-spin PSF (determined from the diameter of the resonance circle) is shifted by $\Delta$z =10 nm, showing that the dark spins lie at or very close to the diamond surface. {\bf c} Dark-spin PSF determination for a second NV center in a vertical (xz) scan. {\bf d} Vertical (xz) NV-MRI of dark spins. Similarly to b, the dark-spin resonance slice image is vertically shifted ($\Delta$z =14 nm for this NV sensor), again suggestive of surface dark spins. Deconvolving the dark-spin image with the PSF gives the spatial distribution of the nearby dark spins, indicating a surface layer above the shallow NV center.}
\end{figure*}

We present imaging experiments mapping the spatial locations of these g=2 dark spins around two separate NV centers by scanning the magnetic tip in three dimensions (Fig \,\ref{Fig3}). Comparing the measured dark-spin PSF to the observed dark-spin resonance slice (Fig. \,\ref{Fig3}b and Fig. \,\ref{Fig3}d, right panel), we find that for both NV centers the dark-spin signal is shifted vertically from the PSF, which shows that the imaged dark spins are located 10 nm and 14 nm above the two NV sensors, respectively (Fig. \,\ref{Fig3}b and Fig. \,\ref{Fig3}d). Given the implantation energy used to form the shallow NV spins, the observed dark-spin location is consistent with them being on the diamond surface (nominal depth of 10 $\pm$ 3 nm).	

An image deconvolution along the xz plane (Fig. \,\ref{Fig3}d) directly shows that the dark spin distribution is spread out in a line, indicating a layer of spins at the diamond surface. This layer of dark spins likely extends further in the xz plane direction, probably with near-uniform coverage over the diamond surface, but laterally distant spins couple more weakly to the single NV and are undetectable when their signal becomes smaller than the measurement noise. As we directly measure the vertical distance between the NV sensor and the dark-spin layer, the density of dark spins in the layer can be found using the dark-spin/NV coupling rate (100 kHz for this NV center)\,\cite{Ofori-Okai2012}. The extracted two-dimensional dark-spin density is 0.5 spins/$nm^2$, which for a surface layer corresponds to a single unpaired electron spin every ~60 surface atoms.

\begin{figure}
\includegraphics[scale=1]{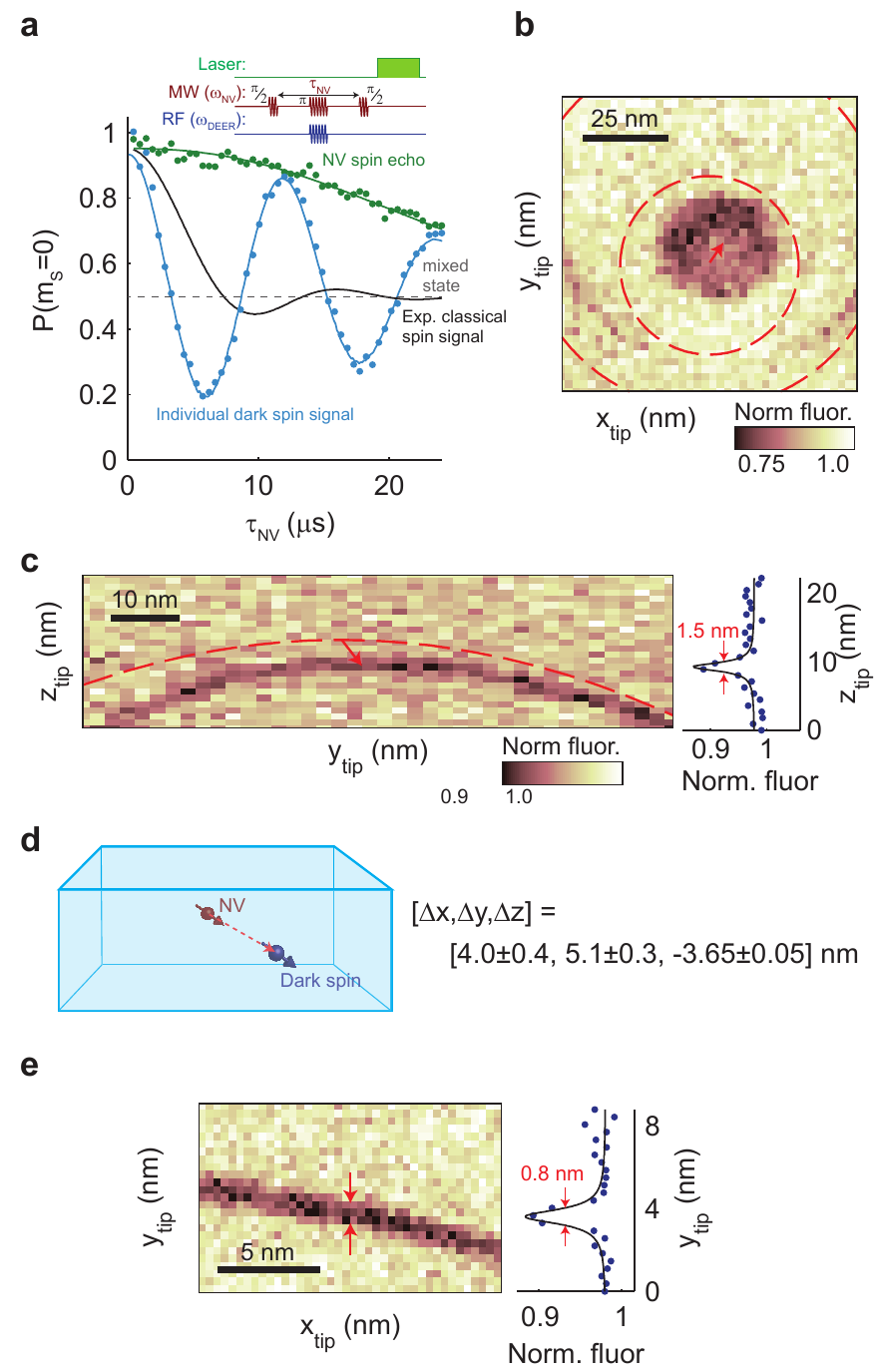}
\caption{\label{Fig4} {\bf a} Coherent dynamics from an individual dark spin strongly coupled to a single shallow NV center. Measured oscillations in the NV DEER signal (pulse sequence shown at top) as a function of spin evolution time $\tau_{NV}$. The DEER signal oscillates coherently and overshoots the NV mixed state, indicated strong coupling of the NV to a single nearby dark electronic spin. For comparison, the expected DEER signal is plotted (black solid line) for comparably strong NV coupling to a single classical spin, demonstrating that the dark-spin effect on the NV DEER signal originates from quantum-projection noise. {\bf b} Lateral (xy) NV-MRI of single, coherently coupled dark spin at 50-nm z-axis tip offset; the dark-spin PSF is illustrated in dashed red lines. The vector connecting the center of the PSF circle and the dark-spin resonance gives the lateral shift ($\Delta$x,$\Delta$y) of the dark spin from the NV. Two distinct resonant slices appear because different sets of dark-spin hyperfine transitions are driven by the applied RF signal. {\bf c} Vertical (yz) NV-MRI of the same coherently coupled dark spin as shown in b. Scanning the magnetic tip across the symmetry plane of the lateral image in b gives the dark-spin vertical shift $\Delta$z, in addition to a second measure of $\Delta$y. The tip-induced magnetic-field gradient along the z-direction provides 1.5 nm spatial resolution. {\bf d} Illustration of the 3D location of the coherently coupled dark spin relative to the NV sensor, as determined from data in b,c. {\bf e} Lateral (xy) NV-MRI of a second coherently coupled dark spin, imaged with 0.8 nm resolution.}
\end{figure}

We also observed that some shallow NV spins are coherently coupled to an individual dark electronic spin, as evidenced by coherent oscillations in the DEER signal as a function of evolution time (Fig. \,\ref{Fig4}a)\,\cite{Shi2013}. We note that the intensity of the observed dark-spin DEER oscillations cannot be explained by a classical single spin, where the spin is modeled by a magnetic moment that can have an arbitrary, continuous magnetization. In that situation, for an unpolarized dark spin, the measured DEER signal would average over all possible magnetizations and the signal would decay to the NV mixed state. However, for a S = 1/2 quantum spin, measurement of its magnetization can only yield two values, giving single-frequency oscillations in the DEER signal, as we observe. This quantum-projection noise enhances the signal for MRI imaging, and in this case it increases the signal-to-noise of dark-spin imaging by 1.9 compared to a classical variance.

Next, we imaged the 3D location of this coherently coupled dark spin by scanning the magnetic tip both laterally (Fig. \,\ref{Fig4}b) and vertically (Fig. \,\ref{Fig4}c) and using the deconvolution method described in Fig. \,\ref{Fig3}. For vertical imaging, we found an offset of the dark-spin position relative to the NV sensor of [$\Delta$y,$\Delta$z] = [5.1 $\pm$ 0.3, -3.65 $\pm$ 0.05] nm, with spatial resolution of 1.5 $\pm$ 0.6 nm given by the spatial width of the dark-spin resonance. For lateral imaging, the dark-spin signal is not only shifted in location, but is also different in size because the dark spin lies at a different depth than the NV sensor and magnetic-field gradients along the z direction are strong compared to the lateral gradients. To achieve high-precision spin-localization in the lateral dimensions, we fit the center of mass of the measured dark-spin PSF and compared this PSF to the center of the response circle, yielding the lateral offset $\Delta$x = 4.0 $\pm$ 0.4 nm. The location of this coherently coupled dark spin relative to the NV sensor is illustrated in the cartoon in Fig. \,\ref{Fig4}d. As an additional example, Fig. \,\ref{Fig4}e displays an xy-plane NV-MRI image of another dark-spin that is coherently coupled to a different NV spin. The lateral spatial resolution in this image is 0.8 $\pm$ 0.4 nm, and we observe that its location is consistent with being potentially on the diamond surface. 

Our NV-MRI demonstration provides the first 3D spatial mapping of dark electronic spins on and near a diamond surface, achieving sub-nanometer resolution. We expect that NV-MRI will be applicable to a wide range of systems in both the physical and life sciences that can be placed on or near the diamond surface and then probed under ambient conditions. For example, one-dimensional spin chains have been proposed as a method for transferring quantum information\,\cite{Yao2011}. A key technical challenge that NV-MRI could address is determining the precise (atomic-scale) location of spins along a chain, which critically influences the fidelity of quantum information transfer. Additionally, individual paramagnetic electron spins with long coherence times at room temperature have attracted interest as potential quantum bits\,\cite{Harneit2002}, but currently such spins cannot be read out individually. NV-MRI would allow for simultaneous control and detection of dark electron spins brought into proximity to NV sensors near the diamond surface. Finally, NV-MRI could image the location of individual electronic spin labels in biological systems, e.g., selectively attached to specific amino acids on a protein\,\cite{Altenbach1990}, which could aid in the determining the structure of proteins.

Furthermore, studying the nanoscale electronic environment on and near diamond surfaces is critical for understanding and maximizing the coherence of NV sensors and quantum bits. We find that the majority of dark spins near shallow NV centers are at the diamond surface, and thus we expect that passivation of the surface to reduce the dark-spin density will improve NV-based sensing and quantum information applications. Alternatively, dark spins at the surface could be initialized with NV-assisted spin-polarization techniques\,\cite{Belthangady2013,Laraoui2013} and then used as a resource for improved sensing: such ancilla sensor spins would effectively amplify magnetic signals\,\cite{Schaffry2011} from samples placed on\,\cite{Mamin2013,Staudacher2013} or scanned\,\cite{Maletinsky2012,Grinolds2013} over the diamond surface. In addition, coherently coupled dark spins, which we identified and imaged can potentially be entangled with the NV sensor to achieve Heisenberg-limited sensing\,\cite{Goldstein2011}, thus dramatically increasing metrology performance.

We gratefully acknowledge S. Kolkowitz, N. de Leon, and C. Bethangady for technical discussions regarding optimizing NV-based sensors and M. Markham and Element Six for providing diamond samples. We would also like to acknowledge fruitful discussions on the detection of dark spins using DEER with M.D. Lukin, A. Sushkov, I. Lovchinsky, N. Chisholm, S. Bennett, and N. Yao. M.S.G. is supported through fellowships from the Department of Defense (NDSEG program) and the National Science Foundation. M.W. is supported through a Marie Curie Fellowship, and K.D.G acknowledges support from the Harvard Quantum Optics Center as an HCOQ postdoctoral fellow. This work was supported by the DARPA QuEST and QuASAR programs and the MURI QuISM.
	

Correspondence and requests for materials should be addressed to A.Y. (\href{mailto:yacoby@physics.harvard.edu}{yacoby@physics.harvard.edu})
\\
\\
\\

\bibliographystyle{apsrev4-1}
\bibliography{nanoMRI}

\end{document}